# Exact Partial Wave Expansion of Optical Beams with Respect to an Arbitrary Origin

A. A. R. Neves,[*] A. Fontes, L. A. Padilha, E. Rodriguez, C. H. Brito Cruz, L. C. Barbosa and C. L. Cesar

Instituto de Física, Universidade Estadual de Campinas, P.O. Box 6165, 13083-970, Campinas, Brazil.

Abstract: Using an analytical expression for an integral involving Bessel and Legendre functions we succeeded to obtain the partial wave decomposition of a general optical beam at an arbitrary location from the origin. We also showed that the solid angle integration will eliminate the radial dependence of the expansion coefficients. The beam shape coefficients obtained are given by an exact expression in terms of single or double integrals. These integrals can be evaluated numerically in a short time scale. We presented the results for the case of linear polarized Gaussian beam.

Partial wave decomposition of incident beams is an important task for any scattering problem of spherical particles and other geometries. For the case of spherical particles the origin of the coordinate must be in the center of the sphere to impose the boundary conditions of continuity of the electric and magnetic fields at the interfaces. Therefore, the partial wave decomposition must be performed for an arbitrary origin and not at any convenient position of high symmetry points of the beam. Although partial wave decomposition is used in many fields, like in quantum mechanic scattering of particles, in optics it requires a full vector description of the electromagnetic fields. This can be a quite complicated problem, especially for highly focused laser beams, where the paraxial limit fails, and all sorts of approximations and tricks have been used to proceed forward and to obtain numerical results [1]. In principle this is an old subject included in text books on Electrodynamics [2] and Mathematical Physics [3], where it is common to express the partial wave expansion of incident optical beams in the form:

$$\mathbf{E}_{inc} = E_0 \sum_{n,m} \left[ \frac{i}{k} G_{nm}^{TM} \nabla \times j_n(kr) \mathbf{X}_{nm}(\theta,\phi) + G_{nm}^{TE} j_n(kr) \mathbf{X}_{nm}(\theta,\phi) \right]$$
$$\mathbf{H}_{inc} = \frac{E_0}{Z} \sum_{n,m} \left[ G_{nm}^{TM} j_n(kr) \mathbf{X}_{nm}(\theta,\phi) - \frac{i}{k} G_{nm}^{TE} \nabla \times j_n(kr) \mathbf{X}_{nm}(\theta,\phi) \right], \quad (1)$$

where $\mathbf{X}_{nm}(\theta,\phi)$ is the vector spherical harmonic, $j_n(kr)$ are spherical Bessel functions and $Z = \sqrt{\mu/\varepsilon}$ is the medium impedance. The incident field has a time varying harmonic component $\exp(-i\omega t)$ that has beam omitted. We will follow the notation given by Jackson [2] where possible. The expansion coefficients $G_{nm}^{TM}$ and $G_{nm}^{TE}$, also called beam shape coefficients BSC [4], are obtained by integrating the radial component of the fields over the solid angle

$$j_n(kr) \begin{bmatrix} G_{nm}^{TM} \\ G_{nm}^{TE} \end{bmatrix} = \frac{kr}{E_0 \sqrt{n(n+1)}} \int Y_{nm}^*(\theta,\phi) \begin{bmatrix} -E_r \\ ZH_r \end{bmatrix} d\Omega, \quad (2)$$

---

[*] Electronic address: aneves@ifi.unicamp.br

For both sides to be mathematically identical a spherical Bessel function must emerge from this solid angle integration. At this point the text books fail, as far as we know, to provide the proof that the integration over the solid angle will actually generate the spherical Bessel function necessary to cancel out the left hand side for an arbitrary optical beam and to obtain a final radial independent expression for the BSC. Sometimes it is suggested to use two radii to obtain the coefficients [2], while others propose to perform integration over the radius to get rid of the radial function on both sides [4]. Either way, the computation of these coefficients are evaluated numerically by using quadratures, finite series technique or localized approximation [5-7] being very time-consuming or, in the last case, an approximation.

We recently showed [8] the analytical expression for the integral:

$$\int_0^\pi d\theta \sin\theta\, P_n^m(\cos\theta)\exp(ikr\cos\alpha\cos\theta)J_m(kr\sin\alpha\sin\theta) = 2i^{n-m}P_n^m(\cos\alpha)j_n(kr), \quad (3)$$

that allowed us to provide an analytical expression for the BSC and to show that the solid angle integration will indeed generate the required spherical Bessel function for an arbitrary normal incident beam, instead of the traditional employed Barton-Davis beam description [9-10]. The BSC are given by an exact, closed expression, without any approximation, in terms of a single, for axially symmetric beams, or double integrals, for non axially symmetric beams. These integrals can be numerically evaluated in a short time scale by common calculation package which also makes it very helpful for numerical evaluation.

To keep the formalism general the electromagnetic fields of an incident arbitrary beam perpendicularly to a lens system will be described by the Angular Spectrum Representation (ASR) [11] in cylindrical coordinates $(\rho,\varphi,z)$, with the origin at the center of the reference sphere of the focusing lens as

$$\mathbf{F}(\rho,\varphi,z) = \frac{ikf\exp(-ikf)}{2\pi}\int_0^{\alpha_{\max}} d\alpha \sin\alpha \exp(ikz\cos\alpha) \\ \int_0^{2\pi} d\beta\, \mathbf{F}_\infty(\alpha,\beta)\exp[ik\rho\sin\alpha\cos(\beta-\varphi)], \quad (4)$$

where the **F** represents either **E** or **B** fields, $\mathbf{F}_\infty(\alpha,\beta)$ is the far field at the reference sphere, and $\alpha$ and $\beta$ are the polar and azimuthal angle respectively subtended by the objective lens aperture ($\alpha_{max} = \sin^{-1}(NA/n_1)$), where NA is the objective numerical aperture and $n_1$ the refractive index of the immersion medium. The far field, separated into perpendicular and parallel components, in terms of the incident beam after an aplanatic lens is given by, [11]

$$\mathbf{F}_\infty(\theta,\phi) = \sqrt{\cos\theta} \left\{ \left[ \begin{pmatrix} F_x \\ F_y \\ 0 \end{pmatrix} \cdot \begin{pmatrix} -\sin\phi \\ \cos\phi \\ 0 \end{pmatrix} \right] \begin{pmatrix} -\sin\phi \\ \cos\phi \\ 0 \end{pmatrix} + \left[ \begin{pmatrix} F_x \\ F_y \\ 0 \end{pmatrix} \cdot \begin{pmatrix} \cos\phi \\ \sin\phi \\ 0 \end{pmatrix} \right] \begin{pmatrix} \cos\phi\cos\theta \\ \sin\phi\cos\theta \\ -\sin\theta \end{pmatrix} \right\}, \quad (5)$$

which shows that the magnetic field coefficients can be extracted from the electric ones with the transformation $F_x \to F_y/Z$ and $F_y \to -F_x/Z$. The Richards and Wolf [12] solution for the fields of high NA objectives are directly recovered from these expressions, but for decomposition purposes it is better to keep the incident beam in the integral form of Eq. (4). To achieve a general solution the first task is to change the cylindrical coordinates $(\rho,\varphi,z)$ for the spherical ones $(r,\theta,\phi)$ with the origin in an arbitrary position by using the transformations $\rho\cos\varphi = r\sin\theta\cos\phi - x_o$, $\rho\sin\varphi = r\sin\theta\sin\phi - y_o$, and $z = r\cos\theta - z_o$, where $(x_o,y_o,z_o)$ is the location of the focal spot with respect to the arbitrary origin. Using these expressions the BSC are given by

$$j_n(kr)\begin{bmatrix}G_{nm}^{TM}\\G_{nm}^{TE}\end{bmatrix} = \mp \frac{ikf\exp(-ikf)}{4\pi E_0} kr\sqrt{\frac{2n+1}{4\pi n(n+1)}\frac{(n-m)!}{(n+m)!}}$$

$$\int_0^{\alpha_{\max}} d\alpha \sin\alpha \sqrt{\cos\alpha} \exp[ik(r\cos\theta - z_o)\cos\alpha] \int_0^{2\pi} d\beta \exp[-ik\rho_o \sin\alpha \cos(\phi_o - \beta)]$$

$$\int_0^\pi d\theta \sin\theta P_n^m(\cos\theta) \int_0^{2\pi} d\phi \exp[ikr\sin\alpha \sin\theta \cos(\phi - \beta)] \exp(-im\phi) \quad , (6)$$

$$\{[F_x(1+\cos\alpha) - (F_x\cos 2\beta + F_y\sin 2\beta)(1-\cos\alpha)]\sin\theta\cos\phi$$
$$+ [F_y(1+\cos\alpha) + (F_y\cos 2\beta - F_x\sin 2\beta)(1-\cos\alpha)]\sin\theta\sin\phi$$
$$- 2(F_x\cos\beta + F_y\sin\beta)\sin\alpha\cos\theta\}$$

With the upper sign and $F = E$, for the $G_{nm}^{TM}$ beam coefficient and the lower sign and $F = H$ for the $G_{nm}^{TE}$ beam coefficient. The first integration in $\phi$ can be solved using,

$$\int_0^{2\pi} d\phi \exp[\pm i x \cos(\phi - \beta)] \begin{bmatrix}\cos\phi\\\sin\phi\\1\end{bmatrix} \exp(-im\phi) =$$

$$-\pi(\pm i)^m \exp(-im\beta) \begin{bmatrix} iJ_{m-1}(x)\exp(i\beta) - iJ_{m+1}(x)\exp(-i\beta) \\ J_{m-1}(x)\exp(i\beta) + J_{m+1}(x)\exp(-i\beta) \\ -2J_m(x) \end{bmatrix} , \quad (7)$$

that leads to, after some algebraic manipulation and the use of Bessel's recurrence relations

$$j_n(kr)\begin{bmatrix}G_{nm}^{TM}\\G_{nm}^{TE}\end{bmatrix} = \pm \frac{ikf\exp(-ikf)}{2E_0} i^m \sqrt{\frac{2n+1}{4\pi n(n+1)}\frac{(n-m)!}{(n+m)!}} \int_0^{\alpha_{\max}} d\alpha \sin\alpha \sqrt{\cos\alpha} \exp(-ikz_0\cos\alpha)$$

$$\int_0^{2\pi} d\beta \exp[-ik\rho_o \sin\alpha \cos(\phi_o - \beta)] \exp(-im\beta)$$

$$\left[-(F_x\sin\beta - F_y\cos\beta)\frac{m}{\sin\alpha}\int_0^\pi d\theta \sin\theta P_n^m(\cos\theta) J_m(kr\sin\alpha\sin\theta)\exp(ikr\cos\theta\cos\alpha)\right.$$

$$\left.+ i(F_x\cos\beta + F_y\sin\beta)\frac{d}{d\alpha}\int_0^\pi d\theta \sin\theta P_n^m(\cos\theta) J_m(kr\sin\alpha\sin\theta)\exp(ikr\cos\theta\cos\alpha)\right] \quad , (8)$$

The $\theta$ integrals are readily performed with the result of Eq. (3) showing the emergence of the required $j_n(kr)$ on the RHS of the equation, providing the cancellation of the spherical Bessel functions, and the BSC final $r$-independent expression in terms of a double integral

$$\begin{bmatrix} G_{nm}^{TM} \\ G_{nm}^{TE} \end{bmatrix} = \pm ikf \exp(-ikf) \frac{i^n}{E_0} \sqrt{\frac{2n+1}{4\pi n(n+1)} \frac{(n-m)!}{(n+m)!}} \int_0^{\alpha_{max}} d\alpha \sin\alpha \sqrt{\cos\alpha} \exp(-ikz_o \cos\alpha)$$

$$\int_0^{2\pi} d\beta \exp[-ik\rho_o \sin\alpha \cos(\phi_o - \beta)] \exp(-im\beta) \quad (9)$$

$$\left[ -(F_x \sin\beta - F_y \cos\beta) \frac{m}{\sin\alpha} P_n^m(\cos\alpha) + i(F_x \cos\beta + F_y \sin\beta) \frac{d}{d\alpha} P_n^m(\cos\alpha) \right]$$

This is a general expression for any type of beam entering the objective aperture. For the very common special case of azimuthally symmetric beams, where $F_{x,y}(\alpha,\beta) = F_{x,y}(\alpha)$, the integration in $\beta$ can also be done and the BSC are expressed only in terms of the single integrals

$$\begin{bmatrix} G_{nm}^{TM} \\ G_{nm}^{TE} \end{bmatrix} = \mp ikf \exp(-ikf) \frac{2\pi i^{n-m} \exp(-im\phi_o)}{E_0} \sqrt{\frac{2n+1}{4\pi n(n+1)} \frac{(n-m)!}{(n+m)!}} \int_0^{\alpha_{max}} d\alpha \sqrt{\cos\alpha} \exp(-ikz_o \cos\alpha)$$

$$\left\{ -\left[ m^2 \frac{J_m(k\rho_o \sin\alpha)}{k\rho_o \sin\alpha} P_n^m(\cos\alpha) - \sin^2\alpha J_m'(k\rho_o \sin\alpha) P_n'^m(\cos\alpha) \right] (F_x \cos\phi_o + F_y \sin\phi_o) \right. \quad (10)$$

$$\left. + im \left[ m J_m'(k\rho_o \sin\alpha) P_n^m(\cos\alpha) - \sin^2\alpha \frac{J_m(k\rho_o \sin\alpha)}{k\rho_o \sin\alpha} P_n'^m(\cos\alpha) \right] (F_y \cos\phi_o - F_x \sin\phi_o) \right\}$$

This result can handle with the same degree of difficulty linearly or circularly polarized beams. An important example of these results is the partial wave decomposition of linearly $x$-polarized TEM$_{0,0}$ incident Gaussian beam, the amplitude becomes,

$$F_x(\alpha,\beta) = E_0 e^{-f^2 \sin^2\alpha / \omega_a^2} \qquad F_y(\alpha,\beta) = 0, \quad (11)$$

The BSC can be easily obtained by numerical evaluation of the single integrals,

$$\begin{bmatrix} G_{nm}^{TM} \\ G_{nm}^{TE} \end{bmatrix} = \pm 2\pi i k f \exp(-ikf) i^{n-m} \exp(-im\phi_o) \sqrt{\frac{2n+1}{4\pi n(n+1)} \frac{(n-m)!}{(n+m)!}}$$

$$\int_0^{\alpha_{max}} d\alpha \sqrt{\cos \alpha} \exp(-f^2 \sin^2 \alpha / \omega_a^2) \exp(-ikz_o \cos \alpha)$$

$$\left\{ \left[ m^2 \frac{J_m(k\rho_o \sin \alpha)}{k\rho_o \sin \alpha} P_n^m(\cos \alpha) - \sin^2 \alpha\, J_m'(k\rho_o \sin \alpha) P_n'^m(\cos \alpha) \right] \cos \phi_o \right.$$

$$\left. + im \left[ m J_m'(k\rho_o \sin \alpha) P_n^m(\cos \alpha) - \sin^2 \alpha \frac{J_m(k\rho_o \sin \alpha)}{k\rho_o \sin \alpha} P_n'^m(\cos \alpha) \right] \sin \phi_o \right\} \quad , \quad (12)$$

For the on axis case, where $\rho_o = 0$, the BSC depends only on the |m| = 1 [13], and the expression (12) is simplified to

$$\begin{bmatrix} G_{n,1}^{TM} \\ G_{n,1}^{TE} \end{bmatrix} = \pm G_n = \pm \pi i k f \exp(-ikf) \frac{i^{n-1}}{n(n+1)} \sqrt{\frac{2n+1}{4\pi}}$$

$$\int_0^{\alpha_{max}} d\alpha \sqrt{\cos \alpha} \exp(-f^2 \sin^2 \alpha / \omega_a^2) \exp(-ikz_o \cos \alpha) \left[ P_n^1(\cos \alpha) - \sin^2 \alpha\, P_n'^1(\cos \alpha) \right] \quad , (13)$$

with an equivalent result for $m = -1$. On analyzing Eq. (13) it can be seen that the pair of BSC are no longer independent and are related. For such a beam the BSC can be rewritten as $G_n$, following standard notations as in Ref. [13]. The other special case is the $z_o = 0$ where little is gained from Eq. (12).

$$\begin{bmatrix} G_{nm}^{TM} \\ G_{nm}^{TE} \end{bmatrix} = \pm 2\pi i k f \exp(-ikf) i^{n-m} \exp(-im\phi_o) \sqrt{\frac{2n+1}{4\pi n(n+1)} \frac{(n-m)!}{(n+m)!}}$$

$$\int_0^{\alpha_{max}} d\alpha \sqrt{\cos \alpha} \exp(-f^2 \sin^2 \alpha / \omega_a^2)$$

$$\left\{ \left[ m^2 \frac{J_m(k\rho_o \sin \alpha)}{k\rho_o \sin \alpha} P_n^m(\cos \alpha) - \sin^2 \alpha\, J_m'(k\rho_o \sin \alpha) P_n'^m(\cos \alpha) \right] \cos \phi_o \right.$$

$$\left. + im \left[ m J_m'(k\rho_o \sin \alpha) P_n^m(\cos \alpha) - \sin^2 \alpha \frac{J_m(k\rho_o \sin \alpha)}{k\rho_o \sin \alpha} P_n'^m(\cos \alpha) \right] \sin \phi_o \right\} \quad , \quad (14)$$

Depending on the values of $n$ and $m$, the BSC would be purely real or imaginary if the beam is placed on the coordinate axis.

We apply this formalism to a Gaussian beam ($\lambda$=800nm and waist $\omega_a$=2.5mm) overfilling an objective aperture of NA = 1.25, and focal length $f$ =1.7mm. By changing the overfilling parameter (beam waist at the objective aperture) one can go from a uniform illumination with diffraction to an immerging Gaussian beam after the objective. Figures 1 and 2 shows the radial component of the ASR and the partial wave reconstructed summed up to $n = 30$ of the electric field for an underfilled objective, while Fig. 3 and 4 shows the results for an overfilled objective aperture. We used the thumb rule $n = k r_o$ to choose the maximum $n$ value. The radial component of electric field of the plots shows how good the results are for different objective overfilling factors. These results were calculated with our code written for Mathematica available at the website http://www.ifi.unicamp.br/foton/cienciasdavida-en.htm. The total time for the $2n+1$ BSC numerical evaluation was 90 seconds of CPU time, and can be actually dropped by almost half by using the relation,

$$I(n,-m) = \frac{(n-m)!}{(n+m)!} I(n,m), \qquad (15)$$

where $I(n,m)$ is the integral in $\alpha$ of Eq. (14).

As a conclusion we developed a new exact and numerically efficient method for the evaluation of BSC for incident optical beams with respect to an arbitrary origin, polarization and amplitude in terms of experimental parameters. The description is presented only in terms of electromagnetic fields instead of Poynting vector or Intensity, making it compatible with electromagnetic vector diffraction problems. We emphasize that the integral (Eq. 3) was fundamental for this development that dramatically simplifies the BSC calculation for an arbitrary translation. We remark that no assumption has been made for the size of the scatterer, thus making it adequate for the most general case of the Mie regime and readily applicable (in the case of optics). The framework of the formalism presented here can also be easily applied to general beams such as Laguerre beams, laser sheets, doughnut and top-hat beams, these types of beams will be deferred to a later study.

This work was partially supported by Fundação de Amparo à Pesquisa do Estado de São Paulo (FAPESP) through the Optics and Photonics Research Center (CePOF). We thank the Coordenação de Aperfeiçoamento de Pessoal de Nível Superior (CAPES) for financial support of this research.

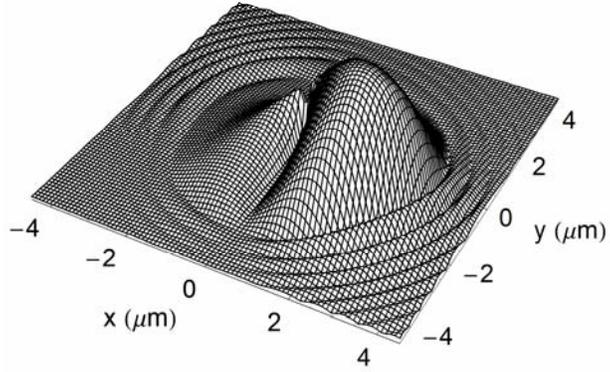

FIG. 1. Plot of the absolute radial electric field reconstructed from partial wave in a plane intersecting $z=0$. Beam centered at point (1.5,-1.5,1.5) μm for the case of an underfilled objective aperture.

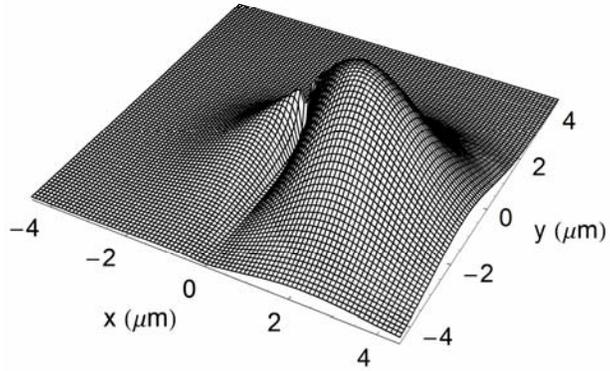

FIG. 2. Plot of the absolute radial electric field from the theoretical description of the ASR in a plane intersecting $z=0$. Beam centered at point (1.5,-1.5,1.5) μm for the case of an underfilled objective aperture.

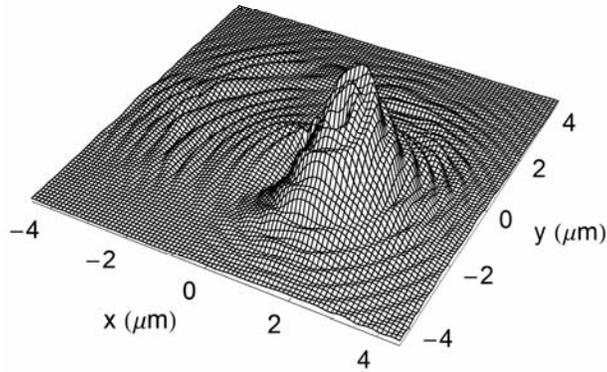

FIG. 3. Plot of the absolute radial electric field reconstructed from partial wave in a plane intersecting $z = 0$. Beam centered at point (1.5,-1.5,1.5) μm for the case of an overfilled objective aperture.

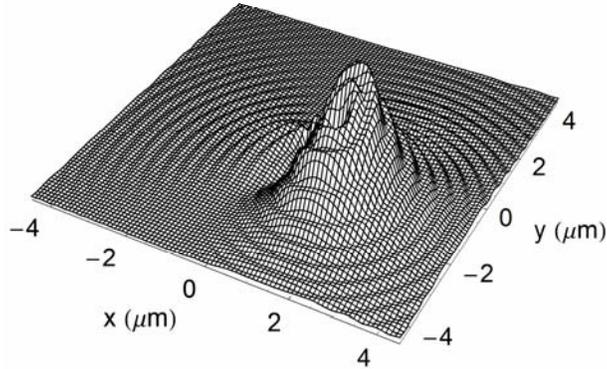

FIG. 4. Plot of the absolute radial electric field from the theoretical description of the ASR in a plane intersecting $z = 0$. Beam centered at point (1.5,-1.5,1.5) μm for the case of an overfilled objective aperture.

11